\documentclass[11pt,dvips]{article}
\usepackage{cite}
\usepackage{epsfig}
\begin{document}

\thispagestyle{empty}
\title{The Poisson ratio of crystalline surfaces} 
\author {Marco Falcioni \hspace{1 cm} Mark J. Bowick\\
{\small Department of Physics, Syracuse University}\\ 
{\small Syracuse, NY 13244-1130, USA}\\ 
{\small \tt http://www.phy.syr.edu/}\\
\and 
Emmanuel Guitter\\
{\small C.E.A.-Saclay, Service de Physique Th\'eorique}\\
{\small F-91191 Gif sur Yvette Cedex, France}\\
\and
Gudmar Thorleifsson\\
{\small Fakult{\"a}t f{\"u}r Physik, Universit{\"a}t Bielefeld}\\ 
{\small D-33615 Bielefeld, Germany}}
\date{SU-4240-649~~Saclay T96/110~~BI-TP 96/46\\[1em] October
1996\\[1em] PACS {\tt 82.65.Dp~~64.60.-i~~87.22.-q}} 
\maketitle

\begin{abstract}
A remarkable theoretical prediction for a crystalline (polymerized)
surface is that its Poisson ratio \(\sigma\) is negative.  Using a
large scale Monte Carlo simulation of a simple model of such surfaces
we show that this is indeed true. The precise numerical value we find
is \(\sigma \simeq -0.32\) on a \(128^2\) lattice at bending rigidity
\(\kappa = 1.1\).  This is in excellent agreement with the prediction
\(\sigma = -1/3\) following from the self-consistent screening
approximation of Le Doussal and Radzihovsky.
\end{abstract}

Crystalline surfaces have been studied extensively in recent years.
Unlike one-dimensional polymers, which are always crumpled, non
self-avoiding crystalline surfaces undergo a continuous phase
transition from a high temperature crumpled phase to a low temperature
flat phase \cite{KN,HW,ADJ}.  The flat phase is characterized by 
long-range orientational order in the normals to the surface.

There are several experimental realizations of crystalline surfaces.
Some, like suspended layers of graphite oxide in aqueous solution
\cite{graphite1,graphite2} or polymerized adsorbed mono-layers, are
chemical systems one can synthetize in the laboratory.  There are also
beautiful biological examples of such surfaces \cite{Skel}: the
cytoskeleton of erythrocytes (red blood cells) is composed of a
network of nodes (actin oligomers) and links (spectrin tetramers).  A
typical skeleton is a triangulated network of roughly 70,000
plaquettes.

The crumpling transition and the flat phase of crystalline surfaces
have been investigated numerically by several authors.  The interested
reader may consult the excellent reviews \cite{DGZJ,jer2}.  Bowick
{\em et}.\ {\em al}.\ have recently performed a large scale simulation
of a triangulated crystalline surface with bending rigidity using free
boundary conditions \cite{BCFTA}.  The largest surface simulated has
32,258 triangles.  The equilibrium distribution is sampled using a
Monte Carlo algorithm with a local Metropolis update.  The action used
has a simple Gaussian potential and a bending energy term:
\begin{equation}
\beta H = \sum_{\langle ij \rangle} (\vec{r}_i - \vec{r}_j)^2 - 
\kappa \sum_{\langle \alpha \beta \rangle} \vec{n}_\alpha 
\cdot \vec{n}_\beta,
\label{model}
\end{equation}
where \(\vec{r}_i\) is the position of node \(i\), \(\vec{n}_\alpha\)
is the unit normal to triangle \(\alpha\) and \(\kappa\) is the
bending rigidity.  For \(\kappa > \kappa_c \simeq 0.79\) the system is
in a {\em flat phase} and it behaves like a membrane with anomalous
elasticity.

The flat phase is characterized by the scaling of the renormalized
effective elastic constants \(\lambda_R \sim \mu_R \sim q^{\eta_u}\)
and of the bending rigidity \(\kappa_R \sim q^{-\eta}\).  The exponent
\(\eta\) can be determined from the scaling of the height-height
correlation function (\(\eta\) is related to the roughness exponent
\(\zeta = 1 - \eta/2\)). The exponent \(\eta_u\) determines the finite
size scaling of the mean square phonon fluctuations. 

We summarize the results of \cite{BCFTA} in Table \ref{table1}.  These
results are compared to the analytical predictions obtained from an
\(\epsilon = 4-D\) expansion (AL) \cite{AL,AGL}, a large-\(d\)
expansion \cite{AGL,GDLP,DG} and a self-consistent screening
approximation (LR) \cite{LDR}.  In this notation \(D\) represents the
dimensionality of the elastic manifold and \(d\) is the dimensionality
of the embedding space (for physical membranes \(D=2\) and \(d=3\).)
Scattering experiments on the red blood cell skeleton give
\(\zeta \simeq 0.65(10)\) \cite{Skel}.

\begin{table*}[t]
\setlength{\tabcolsep}{1.5pc}
\newlength{\digitwidth} \settowidth{\digitwidth}{\rm 0}
\catcode`?=\active \def?{\kern\digitwidth}
\caption{The scaling exponents.}
\label{table1}
\begin{tabular*}{\textwidth}{l@{\extracolsep{\fill}} rrrr} \hline 
& \multicolumn{1}{c}{MC} & \multicolumn{1}{c}{AL} & 
\multicolumn{1}{c}{LR} & \multicolumn{1}{c}{Large-\(d\)} \\ \hline
\(\eta\)   & \(\sim 0.62\) & 24/25 & \(0.81...\) & 2/3 \\
\(\eta_u\) & 0.50(1)       & 2/25  & \(0.36...\) & 2/3 \\
\(\zeta\)  & 0.64(2)       & 13/25 & \(0.59...\) & 2/3 \\
\hline \hline
\end{tabular*}
\end{table*}

In addition to the anomalous scaling of the coupling constants, one of
the most dramatic effects of fluctuations on crystalline surfaces is
the prediction of a {\em negative} Poisson ratio \(\sigma\).  The
Poisson ratio measures the in-plane transverse response of the surface
when stress is applied in the longitudinal direction. It is defined to
be positive for matter which shrinks in the \(\hat{x}\)-direction when
stretched in the \(\hat{y}\)-direction.

Analytical calculations in the context of a self-consistent screening
approximation (LR) \cite{LDR} and an \(\epsilon\) expansion (AL)
\cite{AL} predict that for crystalline surfaces \(\sigma\) is \(-1/3\)
(LR) and \(-1/5\) (AL) respectively.  The unusual sign of the Poisson
ratio is a result of entropic suppression of the height fluctuations
in a membrane under stress \cite{BSS}. This effect is clearly
demonstrated by crumpling a sheet of paper and pulling on two opposite
corners: the sheet expands in the direction transverse to the applied
strain \cite{DRN,CL}.

In this letter we demonstrate numerically that a crystalline surface,
defined by Eq.\ (\ref{model}), indeed has a negative Poisson ratio,
and that it agrees with the prediction from \cite{LDR}.  We compare
our result with previous numerical determinations of \(\sigma\)
\cite{ZDK96}.  We stress that this work extends the analysis of Monte
Carlo data collected in \cite{BCFTA}, and we refer to it for the
details of the simulations.

Consider an asymptotically flat elastic surface at thermal
equilibrium.  In the Monge gauge, its behavior is described by the
partition function
\begin{equation} 
\label{eq:partfun}
Z[\sigma_{ij}] = \int [d r]^3 \exp \left\{ - \int d^2 \varsigma \left[
\frac{1}{2} \left( 2 \mu_0 u^2_{ij} + \lambda_0 u^2_{kk}\right) +
\frac{\kappa_0}{2} (\nabla^2 h)^2 - \sigma_{ij} u_{ij}\right] \right\},
\end{equation}
where \(\varsigma\) is the intrinsic coordinate, \(\mu_0\) and
\(\lambda_0\) are the bare Lam\'e coefficients and \(\kappa_0\) is the
bending rigidity. The stress tensor \(\sigma_{ij}\) represents an
external source linearly coupled to the system.  The strain tensor
\(u_{ij}\), to linear order in \(\vec{u}\), is related to \(\vec{r}\)
by
\begin{eqnarray}
\label{eq:strain}
\vec{r} & = & \vec{s} + \vec{u} + \hat{z} h\\
u_{ij}   & = & \frac{1}{2}(\partial_i u_j + \partial_j u_i + \partial_i
h \partial_j h).
\end{eqnarray}
Here \(\vec{s}\) is the rest (equilibrium) position of the surface,
assumed to lie in the \(x\)--\(y\) plane, and \(h\) is the height of
the surface above the reference plane.

The Poisson ratio can be defined in terms of correlation functions at
zero external stress using linear response theory.  Considering a
diagonal (hydrostatic) stress, the derivation is straightforward and
leads to
\begin{equation}
\label{eq:ourpoiss}
\sigma = - \frac{\langle u_{xx} \, u_{yy} \rangle_c}{\langle u_{yy}^2
\rangle_c}, 
\end{equation}
where the subscript \(c\) indicates the connected part.  Since the
correlation functions are measured in the limit of zero external
stress (\(\sigma_{ij}=0\)) the system is isotropic in the \(x\)--\(y\)
plane and
\begin{equation}
\label{eq:isot}
\langle u_{xx}^2 \rangle_c = \langle u_{yy}^2 \rangle_c.  
\end{equation}

For computational purposes it is convenient to express the strain
tensor in terms of the tangent vectors \(\vec{t}_i = \partial_i
\vec{r}\).  The index \(i\) refers to the intrinsic coordinate system.
The strain tensor is
\begin{equation}
\label{eq:ourstrain}
u_{ij} = \partial_i \vec{r} \cdot \partial_j \vec{r} - \langle
\partial_i \vec{r} \rangle \cdot \langle \partial_j \vec{r} \rangle =
g_{ij} - \delta_{ij},
\end{equation}
where \(g_{ij} = \partial_i \vec{r} \cdot \partial_j \vec{r}\) is the
induced metric, and we have rescaled the intrinsic coordinates so that
\(\langle g_{ij}\rangle = \delta_{ij}\).  Substituting Eq.\
(\ref{eq:ourstrain}) in Eq.\ (\ref{eq:ourpoiss}) we get
\begin{equation}
\sigma = - \frac{\langle g_{xx} \, g_{yy} \rangle_c}{\langle g^2_{yy}
\rangle_c}.
\label{eq:pois2}
\end{equation}
%\begin{figure}
%\centerline{\epsfxsize=2.5in \epsfbox{vectors.eps}}
%\caption{The transformation from hexagonal to orthogonal coordinates.}
%\label{fig:vectors}
%\end{figure}
In terms of a discretized surface the induced metric \(g\)
assumes a simple form.  Take for example the \(xx\) component at point
\(\varsigma\)
\begin{equation}
\label{discrete}
g_{xx}(\varsigma) = \partial_x \vec{r}_{\varsigma} \cdot
\partial_x \vec{r}_{\varsigma} = (\vec{r}_{\varsigma + x} -
\vec{r}_{\varsigma}) \cdot (\vec{r}_{\varsigma + x} -
\vec{r}_{\varsigma}),
\end{equation}
which is the squared length of the ``link'' between the point at
\(\varsigma\) and its neighbor in the \(\hat{x}\) direction. 

The derivation of \(\sigma\) implicitly assumed that the intrinsic
coordinate system is orthogonal.  As our discretized surface is a
triangular lattice, we need to transform the coordinate system.  In
fact, we can directly access only the \(\partial_i
\vec{r}_{\varsigma}\), \(i\) = 1, 2, 3, of Eq.\ (\ref{discrete}), in
the three natural directions of the triangular lattice, while Eq.\
(\ref{eq:pois2}) is expressed in terms of \(\hat{x}\) and \(\hat{y}\).
If $\hat{e}_1$, $\hat{e}_2$, $\hat{e}_3$ are the basis vectors of a
triangular lattice, we define the \(\hat{x}\) direction to overlap
with \(\hat{e}_1\), and we use \(\hat{y} = (\hat{e}_2 + \hat{e}_3) /
\sqrt{3}\).

At this point we can use Eq.\ (\ref{eq:isot}), isotropy, as a
consistency check on our definition of the strain fluctuation.  In
fact we verified that \(\langle g^2_{xx} \rangle_c\), \(\langle g^2_{11}
\rangle_c\), \(\langle g^2_{22} \rangle_c\), \(\langle g^2_{33}
\rangle_c\), and \(\langle g^2_{yy} \rangle_c\) are equal within
numerical accuracy. 

One can find an equivalent definition of \(\sigma\) in terms of
different correlation functions.  Since \(\langle g_{xx} g_{xx}
\rangle_c - \langle g_{xx} g_{yy}\rangle_c = 2 \langle g_{xy}
g_{xy}\rangle_c\) \cite{ZDK96}, it is easy to show that
\begin{equation}
\label{eq:otherpoi}
\sigma = -1 + 2 \frac{\langle g_{xy} g_{xy} \rangle_c}{\langle
g_{xx}^2\rangle_c}.
\end{equation}
We have verified that this definition gives results consistent with
Eq.\ (\ref{eq:pois2}).  But it must be noted that it is more difficult to
use numerical methods to determine \(\sigma\) from Eq.\
(\ref{eq:otherpoi}) as \(\langle g_{xy} \rangle\) is much smaller than
\(\langle g_{xx} \rangle\).

\begin{table*}[t]
\setlength{\tabcolsep}{1.5pc}
\catcode`?=\active \def?{\kern\digitwidth}
\caption{The Poisson Ratio.}
\label{tbl:poisson}
\begin{tabular*}{\textwidth}{l@{\extracolsep{\fill}} r r}
\hline 
\(L\) & \multicolumn{1}{c}{\(\kappa = 1.1\)}&
\multicolumn{1}{c}{\(\kappa = 2.0\)} \\ \hline
32   & \(-0.28(1)\)& \(-0.25(1)\) \\
46   & \(-0.30(2)\)& \(-0.26(2)\) \\
64   & \(-0.30(3)\)& \(-0.28(3)\) \\
128  & \(-0.32(4)\)& \multicolumn{1}{c}{---}  \\ 
 \hline \hline
\end{tabular*}
\end{table*}

\begin{figure}
\centerline{\epsfig{file=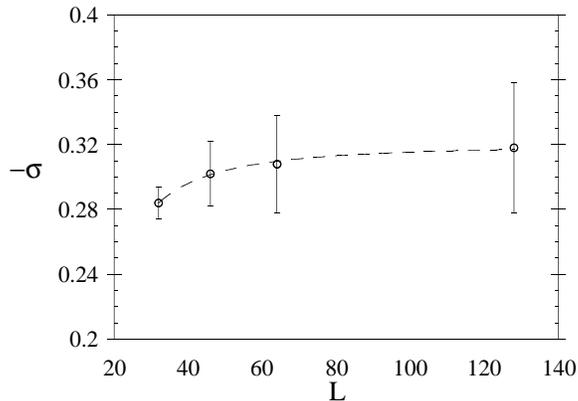, width=3.5in}}
\caption{The Poisson ratio for $\kappa = 1.1$ and
$d = 3L/4$.}
\label{fig:poi}
\end{figure}

We report in Table \ref{tbl:poisson} the measured values of the
Poisson ratio for various sizes of the surface and for two different
values of the bending rigidity \(\kappa\). The errors quoted in the
table are due to statistical fluctuations and are estimated using the
binning method.  In Figure \ref{fig:poi} we plot $\sigma$ for $\kappa
= 1.1$.  We see marked changes in $\sigma$ as the lattice size
increases and as the bending rigidity increases. These discrepancies
are a measure of the uncertainty in our determination of the Poisson
ratio.  We remark that theoretical arguments \cite{AL,LDR} indicate
that the behavior in the whole flat phase is governed by an
infrared-stable fixed point at \(\kappa = \infty\) (the fluctuations
stiffen the surface at long wavelength).  Hence we expect to find the
correct asymptotic behavior as \(L\to\infty\) anywhere above the
crumpling transition.  Larger values of the bending rigidity influence
the results since the auto-correlation times are longer --- it is more
time consuming to gather the equivalent number of independent
configuration.  For both values of \(\kappa\) the results are
consistent with an approach to a value of \(\sigma \sim -0.32\), which
agrees with the theoretical prediction of \(-1/3\) (LR).  The infinite
volume value is extracted from the fit to \(\sigma(L) = \sigma(\infty)
+ {a(\kappa)}/{L^b}\).  The statistical errors on the data do not
allow for a reliable estimate of the exponent \(b\).

Zhang, Davis and Kroll have obtained a different result \(\sigma =
-0.15(1)\) for the Poisson ratio of a tethered membrane in a molecular
dynamics simulation with periodic boundary conditions \cite{ZDK96}.
There are several possible explanations for the discrepancy.  We
stress that the present simulation goes to much larger lattice sizes:
in view of the finite size effects demonstrated in Table
\ref{tbl:poisson}, we believe size to be relevant.  Furthermore, it is
hard to compare the value of the bare parameters used in the two
simulations because both the numerical techniques and the models
studied differ. 

For completeness we mention that Boal, Seifert and Shillcock have
investigated numerically, and with mean field techniques, the Poisson
ratio of 2-dimensional networks under tension \cite{BSS}.  These
models differ from the one we study in that they are strictly planar.
It was found, nonetheless, that entropic effects drive the Poisson
ratio negative.

\begin{figure}
\centerline{\epsfig{file=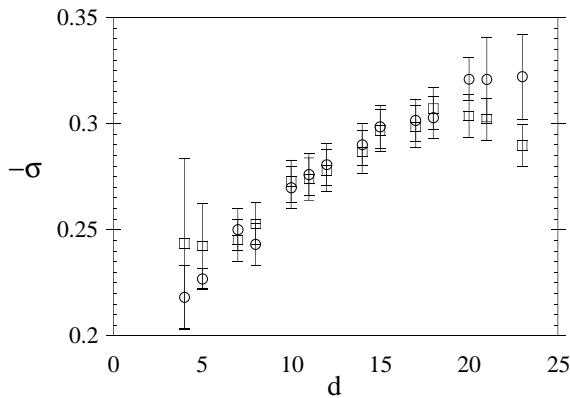, width=3.5in}}
\caption{The Poisson ratio for the 14 subsets at $\kappa = 1.1$ and
$L = 46$. The circles correspond to definition Eq. (\ref{eq:pois2}), while the
squares to definition Eq. (\ref{eq:otherpoi}).}
\label{fig:128}
\end{figure}

A comment now on our treatment of boundary effects.  As noted by
Abraham \cite{A}, large edge fluctuations might influence the results,
even though boundary effects should vanish in the infinite volume
limit. Zhang, Davis and Kroll simulate a surface in a closed cell
\cite{ZDK96}, imposing periodic boundary conditions in the
\(x\)--\(y\) plane only, and, in some cases, dynamically modifying the
cell size.  In our simulations the surface's boundary is free to
fluctuate, and we need to carefully analyze the data in order to
quantify the effect of boundary fluctuations.  To this end, we
restrict the definition of the correlation functions in Eq.\
(\ref{eq:pois2}) to a hexagonal subset of the mesh (the method is
described in great detail in ref.\ \cite{BCFTA}).  The subset excludes
the portions of the surface close to the boundary.  For each size
simulated (\(L=32\), 46, 64, and 128) we construct 14 hexagonal
regions of increasing diameter centered with respect to the bulk
lattice.  We then measure the correlation functions (and hence
\(\sigma\)) restricted to these subsets.  We found that, within our
statistical error, the Poisson ratio for a subset of diameter $d$ is
in fair agreement with the one of an independent surface of size
$L=d$.  Thus we conclude that \(\sigma\) does not suffer much from
boundary effects: presumably these effects cancel out in taking the
ratio of the correlation functions.  The values quoted in
Table~\ref{tbl:poisson} are measured on subsets of diameter
\(d=3L/4\). The justification for this choice of subset size is given
in \cite{BCFTA}.  We show there that the scaling exponents are
insensitive to $d$ for subsets with $d \sim 3L/4$.  Our choice is a
compromise between excluding the boundary and including enough bulk
surface for self-averaging.  In Figure \ref{fig:128} we show the
Poisson ratio for $L=46$ measured on the 14 subsets.

Details of the statistics gathered for the various lattice sizes and
different bending rigidities are reported in \cite{BCFTA}.  For the
largest lattices we ran $74 \times 10^6$ Monte Carlo sweeps
equivalent to 150 independent configurations.  One sweep corresponds
to a Metropolis update of each node of the lattice.

\vspace{1em} 

We acknowledge K.\ Anagnostopoulos, S.\ Catterall, G.\
Jungman, P.\ Le Doussal, D.\ Nelson, and L.\ Radzihovsky for helpful
discussions and suggestions. NPAC has kindly provided computational
facilities.  The research of GT was sponsored by Syracuse University
research funds for part of the work presented here.  The research of
MB and MF was supported by the Department of Energy U.S.A.\ under
contract No.\ DE-FG02-85ER40237. EG thanks the S.U.\ Physics
Department for their kind hospitality.  

\bibliography{paper}

\end{document}